\documentclass[journal=jacsat,manuscript=article]{achemso}
\usepackage{amsmath}
\usepackage{graphicx}
\usepackage{wasysym}
\usepackage{listings}
\usepackage{color}
\usepackage[T1]{fontenc}
\newcommand{\onlinecite}[1]{\hspace{-1 ex} \nocite{#1}\citenum{#1}}

\title{AlCl$_{3}$-dosed Si(100)-2$\times$1: Adsorbates, chlorinated Al chains, and incorporated Al}

\author{Matthew S. Radue}
\affiliation{Laboratory for Physical Sciences, 8050 Greenmead Drive, College Park, MD 20740, USA.}
\author{Sungha Baek}
\affiliation{Department of Physics, University of Maryland, College Park, MD 20742, USA.}
\author{Azadeh Farzaneh}
\affiliation{Department of Materials Science and Engineering, University of Maryland, College Park, MD 20742, USA.}
\author{K.J. Dwyer}
\affiliation{Department of Physics, University of Maryland, College Park, MD 20742, USA.}
\author{Quinn Campbell}
\affiliation{Center for Computing Research, Sandia National Labortaries, Albuquerque, New Mexico 87185, USA.}
\author{Andrew D. Baczewski}
\affiliation{Center for Computing Research, Sandia National Labortaries, Albuquerque, New Mexico 87185, USA.}
\author{Ezra Bussmann}
\affiliation{Sandia National Laboratories, Albuquerque, New Mexico 87185, USA.}
\author{George T. Wang}
\affiliation{Sandia National Laboratories, Albuquerque, New Mexico 87185, USA.}
\author{Yifei Mo}
\affiliation{Department of Materials Science and Engineering, University of Maryland, College Park, MD 20742, USA.}
\author{Shashank Misra}
\affiliation{Sandia National Laboratories, Albuquerque, New Mexico 87185, USA.}
\author{R.E. Butera}
\email{rbutera@lps.umd.edu}
\affiliation{Laboratory for Physical Sciences, 8050 Greenmead Drive, College Park, MD 20740, USA.}
\date{\today}

\begin{document}

\begin{tocentry}
\vfill
\includegraphics[width=\textwidth]{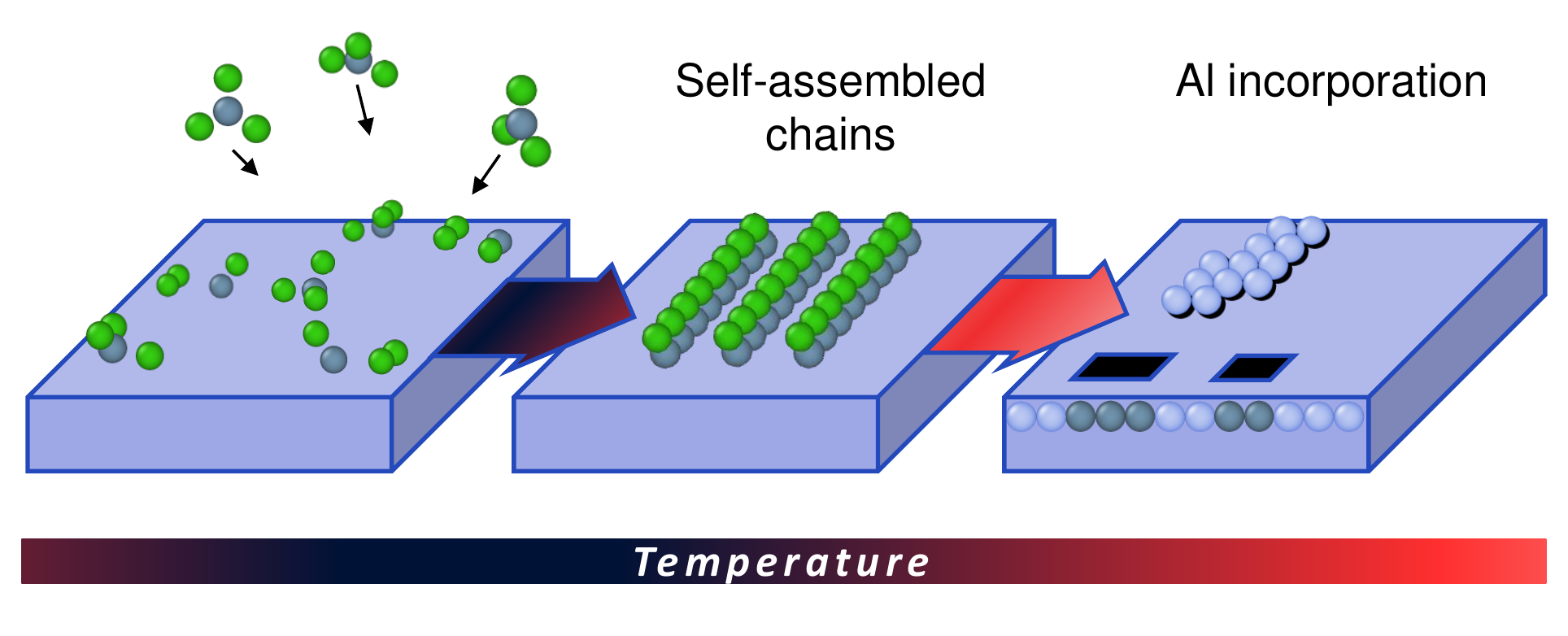}
\vfill
\end{tocentry}

\maketitle

\pagebreak

\begin{abstract}
The adsorption of AlCl$_{3}$ on Si(100) and the effect of annealing the AlCl$_{3}$-dosed substrate was studied to reveal key surface processes for the development of atomic-precision acceptor-doping techniques. This investigation was performed via scanning tunneling microscopy (STM), X-ray photoelectron spectroscopy (XPS), and density functional theory (DFT) calculations. At room temperature, AlCl$_{3}$ readily adsorbed to the Si substrate dimers and dissociated to form a variety of species. Annealing of the AlCl$_{3}$-dosed substrate at temperatures below 450 $^{\circ}$C produced unique chlorinated aluminum chains (CACs) elongated along the Si(100) dimer row direction. An atomic model for the chains is proposed with supporting DFT calculations. Al was incorporated into the Si substrate upon annealing at 450 $^{\circ}$C and above, and Cl desorption was observed for temperatures beyond 450 $^{\circ}$C. Al-incorporated samples were encapsulated in Si and characterized by secondary ion mass spectrometry (SIMS) depth profiling to quantify the Al atom concentration, which was found to be in excess of 10$^{20}$ cm$^{-3}$  across a $\sim$2.7 nm thick $\delta$-doped region. The Al concentration achieved here and the processing parameters utilized promote AlCl$_{3}$ as a viable gaseous precursor for novel acceptor-doped Si materials and devices for quantum computing.

\end{abstract}

\pagebreak

\section{Introduction} \label{introduction}

The field of atomic precision advanced manufacturing (APAM) \cite{APAM} has been steadily advancing such that P-doped, Si-based devices are fabricated, manipulated, and measured in the lab with increasing ease.\cite{SNL:Optical, NIST:Optical, Reza:AC_Litho, SNL:Surface_Gate, SNL:Ellipsometry, UNSW:RF_Readout, UNSW:3D_Readout} The surface chemical processes associated with PH\textsubscript{3} adsorption on Si(100) have been thoroughly studied \cite{Warschkow_PH3:2016} and extensively leveraged to fabricate spin qubit structures with APAM techniques.\cite{Kane:1998, UNSW:2_qubit} Only recently has there been progress in extending capabilities to include acceptor dopants that would enable the realization of hole-based qubits \cite{Salfi:2016, Salfi_PRL:2016} and complex device structures.\cite{Fuhrer:diborane, Curson:AsH3} Of particular note, the first demonstration of APAM-fabricated, acceptor-doped devices (an atomic-scale wire, tunnel junction, and an atomically abrupt pn-junction) were recently demonstrated utilizing B\textsubscript{2}H\textsubscript{6} as the dopant precursor.\cite{Fuhrer:diborane} However, these devices required activation annealing upwards of 850 $^{\circ}$C to achieve a sheet resistance of $\sim$1 k$\Omega$, prompting further investigation to find a more ideal precursor.\cite{campbell2021}

The identification of an APAM-compatible acceptor precursor has been of recent interest due to the proposal from Shim and Tahan on the potential realization of atomic-scale superconducting devices in Si.\cite{Shim:2014} Combining the superconducting properties of ultra-doped silicon with the ability to lithographically pattern atomic-scale device structures and fully encapsulate the device in epitaxial silicon is seen as a potential pathway to significantly reduce interfacial dielectric loss mechanisms in current superconducting qubit devices.\cite{Shim:SSQD} Superconductivity in B-doped silicon has already been demonstrated on larger scale devices through gas immersion laser doping.\cite{CAMMILLERI200875} However, Al doping is expected to produce an order-of-magnitude higher critical temperature than B doping.\cite{bourgeois2007} Recent investigations have attempted to utilize $\delta$-doping from evaporated aluminum on Si(100) to achieve superconductivity with no clear signs of a superconducting transition.\cite{NIST:Si_Al} The realization of the device structures proposed by Shim and Tahan requires area selective deposition of the chosen dopant precursor and while the adsorption of AlH\textsubscript{3} on Si(100) was explored theoretically,\cite{Smith_2017} it cannot currently be readily produced for dosing experiments.  With recent demonstration of atomic-scale patterning of Cl-Si(100) \cite{ClLitho} and the use of BCl\textsubscript{3} in superconducting structures, the exploration of a Cl-based Al precursor, AlCl\textsubscript{3}, is within reason. Both BCl$_{3}$ and AlCl$_{3}$ are readily available and used extensively in atomic layer deposition applications.\cite{ferguson2002,pilli2018,rontu2018,yun1997} Although investigations have explored the adsorption of AlCl\textsubscript{3} on Si(111) \cite{uesugi1993,uesugi1994,takiguchi1994} and Ge(100),\cite{Ge:AlCl3} a thorough investigation of the atomic-level details of the reaction with Si(100) has not yet been performed.

In this paper we investigate the adsorption of AlCl$_{3}$ on Si(100) using a scanning tunneling microscope (STM) to monitor the evolution of the surface as a function of temperature as Al atoms are incorporated into the Si substrate. We find AlCl$_{3}$ adsorbs readily at room temperature forming cyclic structures and dissociated species. We observe the formation of self-assembled chlorinated aluminum chains (CACs) not previously observed on Al-dosed substrates, prompting an atomic model evaluated by density functional theory (DFT) calculations to identify the composition of the chains. According to this model, the presence of Cl allows Al to break Si dimer bonds by stabilizing the dimer-breaking Al atoms. Further annealing leads to the breakup of the CACs, incorporation of Al, and subsequent desorption of Cl. The STM evidence of these steps is identified and discussed herein. While there are further advancements to be made to realize atomic-precision acceptor-doping with Al, the behavior and thermal stimulation of the relevant precursors on Si(100) must be thoroughly understood if atomic manipulation is to be successful. This study serves to advance the topic in the case of AlCl$_{3}$, and thereupon we find AlCl$_{3}$ to be a favorable precursor.

The experimental and computational methods are described in section II, the results are presented and discussed in section III, and the main conclusions are summarized in section IV. The results section is subdivided as follows: the initial adsorption of AlCl\textsubscript{3} (section III A), the effect of annealing the AlCl\textsubscript{3}-dosed surface leading to the formation of CACs (section III B), the atomic structure of the CACs (section III C), the incorporation of Al (section III D), and the desorption of Cl (section III E).

\section{Methods} \label{methods}
\subsection{Experimental Methods}

The experiments were performed in an ultra-high vacuum (UHV) system with base pressure $P$ \mbox{$< 2.7 \times 10^{-9}$ Pa} ($2.0 \times 10^{-11}$ Torr) using a ScientaOmicron VT-STM with a ZyVector STM Lithography control system. The Si wafers used in these experiments were p-type, obtained from ITME. They were B-doped to $\rho = $\mbox{0.07 $\Omega\cdot$cm} to \mbox{0.09 $\Omega\cdot$cm}, and oriented within 0.5$^{\circ}$ of (100). The Si samples were cut to 4 mm $\times$ 12 mm in size and cleaned by sonication in acetone, methanol, and isopropanol. They were then immediately mounted on a ScientaOmicron XA sample plate, and loaded into the UHV chamber. Clean Si(100)-(2$\times$1) surfaces were prepared by flash annealing the sample to 1225 $^{\circ}$C following the procedure found in Ref. \onlinecite{Trenhaile_O:2006}. Samples were scanned in the STM to check for contaminants and then re-flashed before dosing.

AlCl$_{3}$ dosing was done with the sample at room temperature, but performed shortly after clean surface preparation to minimize water contamination. A sample cylinder containing 40 g of anhydrous AlCl$_{3}$ powder packed in Ar (purity $>$0.9999, obtained from Strem Chemicals) was heated to $\sim$170 $^{\circ}$C to facilitate sublimation and the gaseous AlCl$_{3}$ was carried with Ar into the UHV chamber via a precision leak valve. A residual gas analyzer (RGA) in the dosing chamber showed the presence of mainly Ar, H$_{2}$, HCl, N$_{2}$, and some water in addition to various AlCl species during dosing. While some HCl generation is inevitable, Ar was left in the bottle to prevent the generation of excess HCl. The AlCl$_{3}$ doses given here are expressed in units of Langmuir (1 L = 10$^{-6}$ Torr$\cdot$s). The reported values were estimated based on the partial pressure of the AlCl$_{3}$ in Ar calculated from the vapor pressure at $\sim$170 $^{\circ}$C. These dose values are experimental parameters with a relative uncertainty of $\sim$0.1 when comparing different doses; however, they should be taken as likely upper bounds on the true dose with large absolute uncertainties due to the difficulty of determining an accurate partial pressure. AlCl\textsubscript{3} dosing was performed with the sample at room temperature and the sample temperature during annealing was measured with an optical pyrometer.

X-ray photoelectron spectroscopy (XPS) spectra of AlCl$_{3}$-dosed samples were collected to monitor Cl content on the samples using a Staib Instruments DESA 100 cylindrical mirror analyzer. A Mg anode x-ray source was used for measurements, also from Staib. After XPS and STM analysis, samples were encapsulated in $\sim$35 nm Si and measured using secondary ion mass spectrometry (SIMS) depth profiling (performed externally by EAG Materials Science) to determine the Al $\delta$-layer concentration. Si was deposited at a rate of $\sim$0.01 nm/s using a MBE Komponenten SUSI-63 sublimation source. Samples were not heated during deposition, although thermal radiation from the deposition source heated the sample to $\sim$245 $^{\circ}$C, as measured by a nearby thermocouple.

\subsection{Computational Methods}

Two separate modeling setups were used to obtain the DFT results herein. Both approaches are described in detail in the Supporting Info document, and their contributions to the results are specified. In order to ensure the modeling approaches agree, the adsorption energies for the AlCl\textsubscript{3} adsorbate structures were computed via both methods and found to be in good agreement (see Table 1 in Supporting Info).

For reported adsorption energies of the initial AlCl\textsubscript{3} adsorbates, the following equation was used:
\begin{equation}
E_{\rm a} = E_{\rm slab/adsorbate} - E_{\rm slab} - E_{\rm AlCl_3},
\end{equation}
\sloppy where $E_{\rm a}$ is the adsorption energy of the molecule on the silicon surface, $E_{\rm slab/adsorbate} $ is the total energy of the adsorbate on the slab, $E_{\rm slab} $ is the total energy of the slab without any adsorbate, and $E_{\rm AlCl_3}$ is the total energy of the isolated AlCl$_{3}$ molecule. Negative values of $E_{\rm a}$  imply a thermodynamically favorable adsorption energy for that configuration.

For the chain models we will consider, the Al-to-Cl ratio is greater than 1:3, the Al-to-Cl ratio of the precursor. To compute the adsorption energy, the excess Cl is assumed to be bound to Si as Cl-Si-Si-Cl (dimer with two monochlorinated Si). For a chain model with $N_{\rm Al}$ Al atoms (supplied by $N_{\rm Al}$ AlCl\textsubscript{3} molecules) and $N_{\rm Cl}$ Cl atoms, we have $3N_{\rm Al}-N_{\rm Cl}$ remaining Cl atoms not included in the chain. The adsorption energy of the chain and remaining Cl is computed as,
\begin{multline} \label{eq:chain}
E_{\rm a} = (E_{\rm slab_1/chain} - E_{\rm slab_1}) + \frac{1}{2} (3N_{\rm Al}-N_{\rm Cl}) (E_{\rm slab_2/Cl_2} - E_{\rm slab_2}) - N_{\rm Al} E_{\rm AlCl_3}
\end{multline}
\sloppy where $E_{\rm slab_1/chain}- E_{\rm slab_1}$ is the difference between the total energies of the chain on a slab and the slab by itself, $E_{\rm slab_2/Cl_2} - E_{\rm slab_2}$ is the difference between the total energies of an adsorbed Cl\textsubscript{2} molecule on a slab and that slab alone, and $E_{\rm AlCl_3}$ is the total energy of the isolated AlCl\textsubscript{3} molecule. To compare chains requiring differing amounts of AlCl\textsubscript{3}, we divide the chain adsorption energy by the number of AlCl\textsubscript{3} molecules needed, giving an average adsorption energy.

Simulated STM images were produced according to the Tersoff-Hamann approximation,\cite{tersoff1985} where the tunneling current is proportional to the surface local density of states. The local density of states was computed by the band-decomposed charge density for energies just below or above Fermi energy. From this, we select an isosurface of constant charge density ($1\times10^{-5}$ \textit{e}\AA$^{-3}$) and follow its variation in height to replicate the constant-current scanning mode. For each image, the energy range is found from the reported voltage. To produce symmetric Si dimer images, the images of opposite buckling patterns were averaged together and smoothed by Gaussian smoothing.

\section{Results and Discussion} \label{results}
\subsection{AlCl\textsubscript{3} Adsorption} \label{ssA}

\begin{figure*}
\centering
\includegraphics[width=\textwidth]{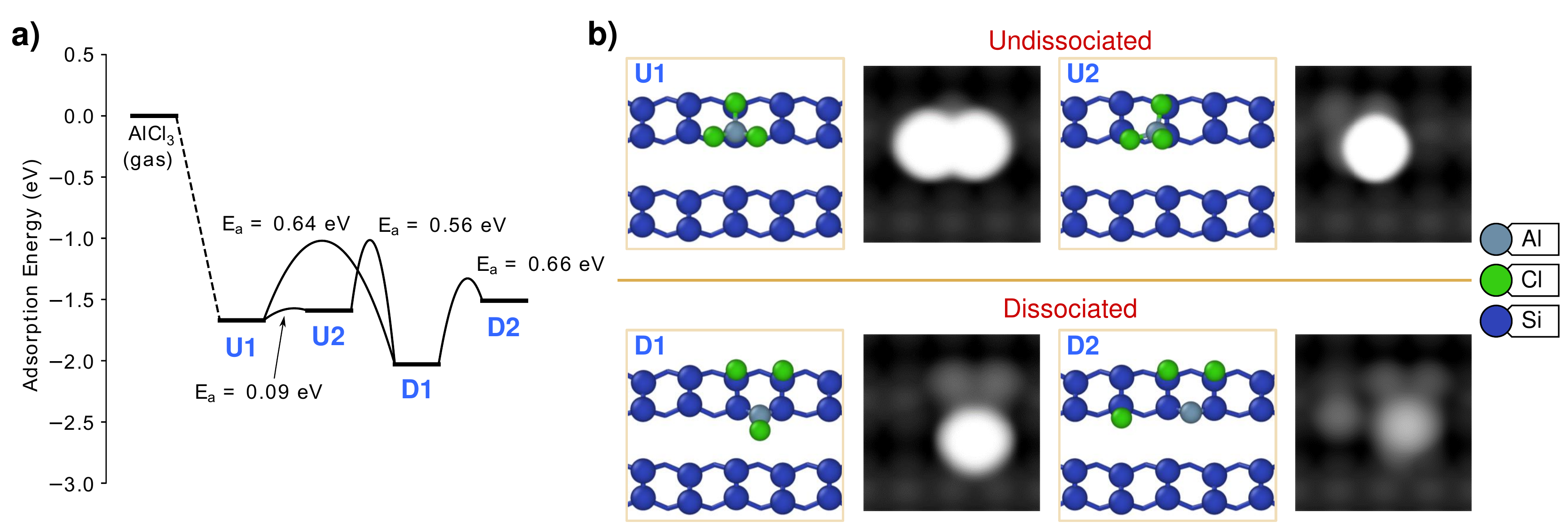}\hfill%
\caption{(a) The adsorption energies of key AlCl\textsubscript{3} and dissociated adsorbates and the reaction barriers for converting between configurations. (b) The DFT-optimized structures for the configurations in (a) and the corresponding simulated filled-state (-1.6 V) STM images. Upon interaction with the Si surface, it is energetically favorable for two Cl atoms to dissociate, leaving a bridging AlCl unit. The atomic color legend for this figure (and following figures) is shown on the far right}%
\label{dft}%
\end{figure*}

We first present our theoretical work on AlCl\textsubscript{3} to gain useful atomic-level insight for interpreting the following experimental images. The DFT-calculated adsorption and dissociation pathway of AlCl\textsubscript{3} on Si(100) is shown in Fig.~\ref{dft}(a). Four critical adsorbate structures were found along the pathway, which are depicted in Fig.~\ref{dft}(b). Included in panel (a) are the adsorption energies of the adsorbates and the energy barriers for transitioning between structures. The results reveal that adsorption proceeds without barrier into an undissociated structure (U1), forming a four-membered ring with a Si dimer, Si-Al-Cl-Si. (See Supporting Information for an additional view of the structure.) This structure has been discovered previously, and our results are in good agreement.\cite{ferguson2009} With low barrier, another Cl can also bond with Si, as seen in the U2 structure. Since the barrier for this reaction is only 0.09 eV, rapid pivoting between the U1 and U2 structures is possible. Dissociation of one Cl did not appear to be advantageous, since no immediate AlCl\textsubscript{2} structures were found. Rather, two Cls can simultaneously dissociate from either the U1 or U2 structures with energy barriers 0.64 eV and 0.56 eV, respectively. The resulting structure (D1) consists of an AlCl unit after two Cls are dissociated. The D1 structure was the most favorable adsorbate structure found. Moreover, a third Cl can also dissociate leaving a lone bridging Al, as in D2. Note that variations are possible in the location of the dissociated Cls with respect to Al, such as Cl dissociating across the dimer row.

\begin{figure}
\centering
\includegraphics[width=0.5\textwidth]{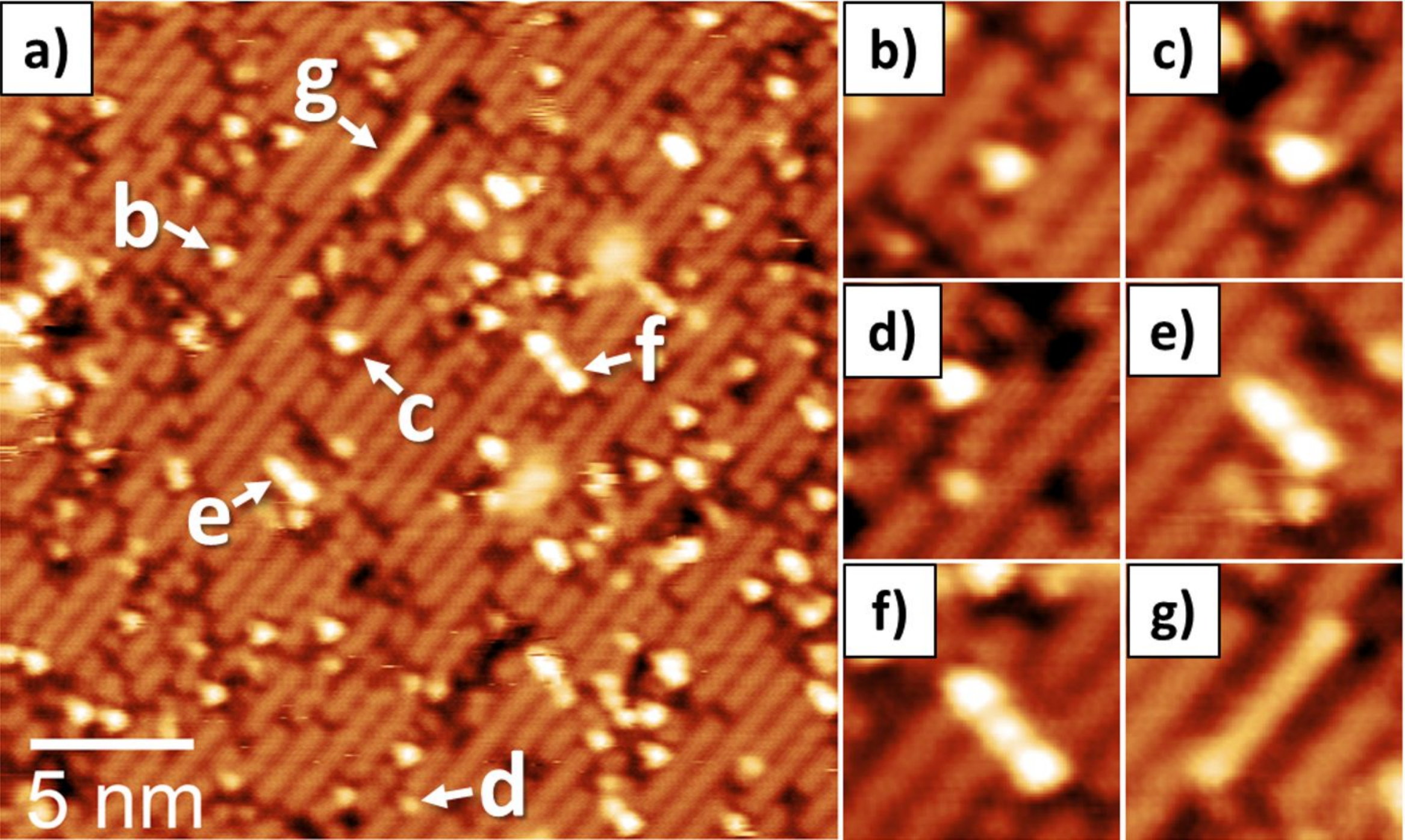}\hfill%
\caption{Low-dosed sample, no annealing. (a) Filled-state STM image (-1.6 V, 0.2 nA) of a Si(100) surface following exposure to AlCl\textsubscript{3} (19 L dose). (b) through (g) show magnified views of several Al-related features visible in (a). Feature (b) is likely an undissociated AlCl\textsubscript{3}. Feature (c) is proposed to be a bridging AlCl unit. Feature (d) is proposed to be an Al with no Cl atoms directly bonded it. (e) and (f) show short Al chains. (g) An elongated Al-Cl feature running along the dimer row.}%
\label{adsorption}%
\end{figure}

Fig.~\ref{adsorption} shows a filled-state STM image of a Si(100) surface exposed to AlCl\textsubscript{3} at a low-coverage dose of \mbox{$\sim$19 L}. As stated earlier, the actual dose is likely to be lower than the experimental value with the expected AlCl$_{3}$ sticking coefficient being near unity for barrierless adsorption. Upon exposure to AlCl\textsubscript{3}, we observed several Al-related features on the surface. Panels (b) through (g) show magnified views of several notable features visible in (a). While there are other gaseous constituents in the chamber during dosing that could contribute to these features, those are predominantly Ar and H$_{2}$ (low sticking coefficient) as well as HCl, which does not appear to be present in significant amounts on our dosed surfaces. HCl appears in STM as small, alternating round features\cite{hsieh2010} that we have observed when the Ar is removed from the AlCl$_{3}$ bottle before dosing (see Supporting Information Figure 2). However, as stated before, Ar was left in the bottle for these experiments, mitigating HCl generation.

We can directly compare the observed species with the simulated filled-state (-1.6 V) STM images computed for the key modeled adsorbates, shown in Fig.~\ref{dft}(b). The undissociated adsorbed AlCl\textsubscript{3} images as either one or two bright spots on the dimer row and offset from the center of the row. The low-energy barrier between U1 and U2 suggests that it could image as one combined feature. In this case, a blended feature would remain on the row, offset from the center. Consider that the U1 configuration encroaches on the neighboring dimers on both sides of the Al-bonded dimer; whereas, the U2 configuration is contained efficiently on exactly two dimers. Therefore, it is also possible that in the presence of neighboring dimer termination the adsorbate might become pinned in the U2 configuration. Feature (b) is assigned to the undissociated AlCl\textsubscript{3} adsorbate on the basis of the location of the bright feature. In contrast, the bridging AlCl in D1 appears as a bright spot about halfway between diagonally adjacent dimers and so falls in between dimer rows. Feature (c) shows a bright feature halfway between dimer rows, which we match to the bridging AlCl unit. When all Cls are dissociated, the lone Al is shown to image noticeably dimmer. This dimmer feature is not exactly halfway between rows, but is closer to the row the Al is bonded to. This matches well with feature (d). Other interesting features are also marked. Fig.~\ref{adsorption}(e) and (f) show what appear to be short Al chains, which are known to run perpendicular to the dimer row direction.\cite{brocks1993,itoh1993,kotlyar2002}  An elongated feature running along a dimer row is shown in (g). This feature foreshadows the CAC, which we will examine in the following sections. Based on the analysis in Section C, the (g) feature is likely a chain of alternating Al and Cl atoms.

Due to the proclivity of AlCl\textsubscript{3} to dimerize, we also consider the adsorption of Al\textsubscript{2}Cl\textsubscript{6}. Our DFT results on the adsorption and dissociation of Al\textsubscript{2}Cl\textsubscript{6} are found in the Supporting Information document. Initial adsorbate structures were identified, and simulated STM images were generated from these adsorbates. However, we do not find the identified Al\textsubscript{2}Cl\textsubscript{6} adsorbates to be prevalent in Fig.~\ref{adsorption}. Instead, the computed AlCl\textsubscript{3} adsorbates seemingly best agree with the features in Fig.~\ref{adsorption}, as there is an abundance of singlet features, like (b), (c), and (d). We speculate that the presence of Ar may be a factor in reducing dimerization, since it is also observed to prevent HCl formation. That being said, a mixture of species ought to be expected, and Al\textsubscript{2}Cl\textsubscript{6} should be present in some amount. While we cannot rule out Al\textsubscript{2}Cl\textsubscript{6}, the results lead us to suspect that monomeric species are very common on the surface.

\subsection{Effect of Annealing and Observation of Unique Chains} \label{ssB}

\begin{figure*}
\centering
\includegraphics[scale=0.8]{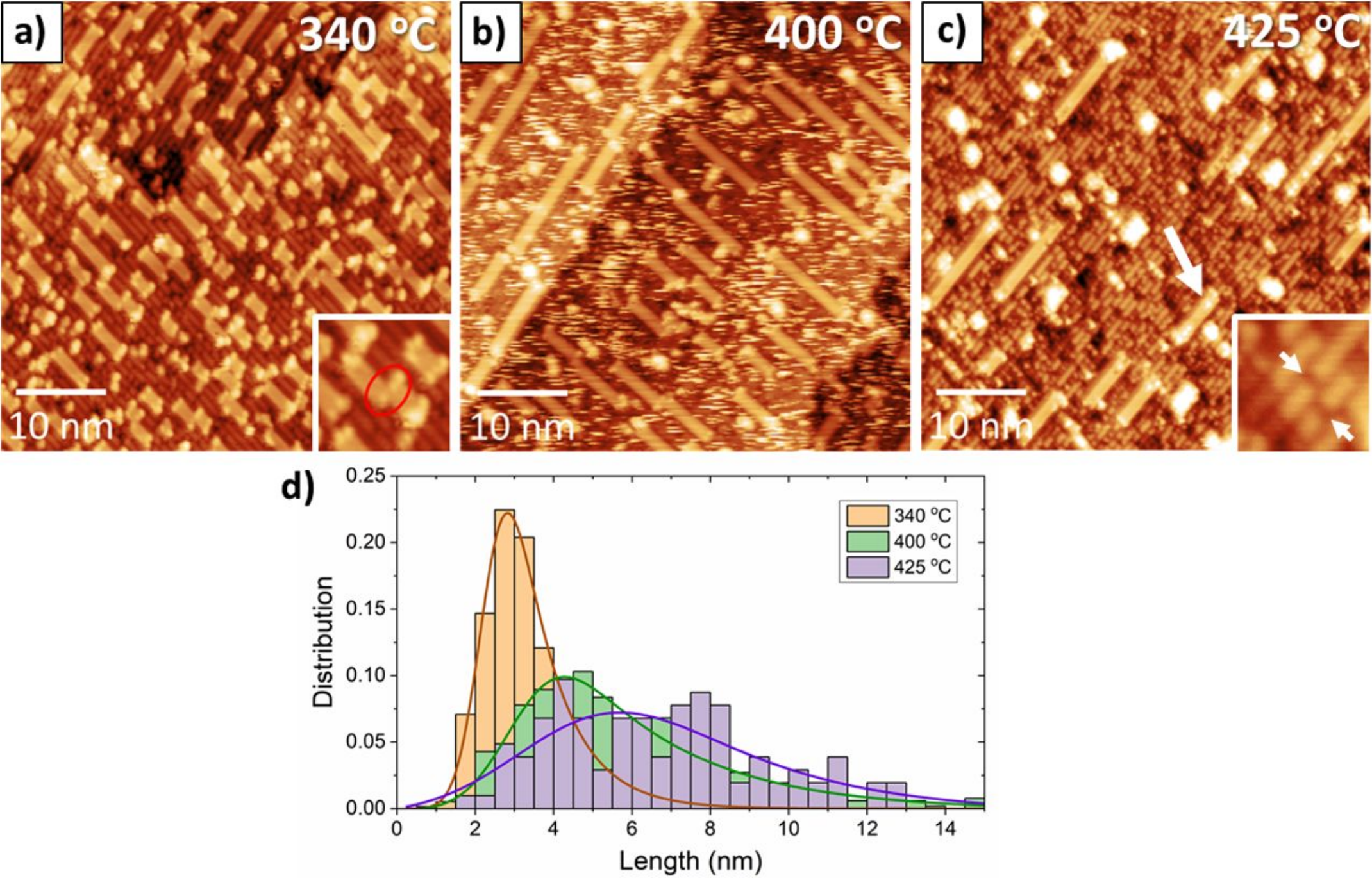}\hfill%
\caption{High-dosed samples, low-temperature annealing. Representative filled-state STM images (a: -1.7 V, 0.15 nA, b: -1.85 V, 0.06 nA, c: -1.8 V, 0.16 nA) of Si(100) dosed with 81 L of AlCl\textsubscript{3} after annealing for 5 minutes to form chlorinated Al chains at (a) 340 $^{\circ}$C, (b) 400 $^{\circ}$C, and (c) 425 $^{\circ}$C. CACs can be seen running parallel to the Si(100) dimer rows. (d) Shows the distribution of chain lengths as a percentage and indicates that the distribution of chain lengths broadens and increases for higher annealing temperatures. The length distributions were fit with an exponentially modified Gaussian shown as solid lines.}%
\label{AlCl3_low_anneal}%
\end{figure*}

Figure~\ref{AlCl3_low_anneal} shows filled-state STM images of Si(100) after exposure to 81 L of AlCl$_{3}$ and annealing at (a) 340 $^{\circ}$C, (b) 400 $^{\circ}$C, and (c) 425 $^{\circ}$C. Annealing at 340 $^{\circ}$C led to the formation of short chains elongated along the dimer row direction, as shown in Fig.~\ref{AlCl3_low_anneal}(a). These chains are unique to this system and differ from other Al-related chains\cite{brocks1993,itoh1993,kotlyar2002} in that they are oriented parallel to the dimer row direction instead of perpendicular. Significantly, every chain observed had a width of $\sim$1.6 nm. Occasionally, neighboring chains would meet at the ends, as highlighted by the circle in Fig.~\ref{AlCl3_low_anneal}(a), so that at these locations the width was $\sim$2.4 nm. This is less than twice the width of a single chain, indicating some overlap of the chain structures. In general, neighboring chains maintained a lateral spacing of at least one atomic site. A step runs through the upper portion of the image where the chains are observed to have rotated their orientation direction by 90$^{\circ}$ on the different terraces as do the dimer rows. Few other surface species can be found on the terrace, which were limited to Si vacancies and c-type defects. Of particular note, the area between the chains appears to be nearly free of adsorbed Cl which would image as a darkened Si dimer.

Annealing at 340 $^{\circ}$C for 5 minutes, produced an average chain density of 4.01(30) $\times$ 10$^{4}$ $\mu$m$^{-2}$. The average chain length was 3.33(5) nm. However, annealing at 400 $^{\circ}$C, we find the average chain length increased to 5.9(1) nm, and the chain density decreased to 1.713(83) $\times$ 10$^{4}$ $\mu$m$^{-2}$.  As discussed in Part E (see Fig.~\ref{XPS}), the total surface coverage of the chains remained nearly unchanged. The area between the chains has become quite noisy, but the chains have a normal appearance. Annealing at 425 $^{\circ}$C leads to a noticeable change in the chain density along with a significant drop in the overall surface coverage. At this temperature, the average chain density was 6.44(67) $\times$ 10$^{3}$ $\mu$m$^{-2}$. The average chain length has increased further to 6.7(3) nm, but as seen in Fig.~\ref{AlCl3_low_anneal}(c) we now observe additional features on the surface that appear to be adsorbed Cl, i.e. Cl-terminated Si dimers, as marked in the inset of Fig.~\ref{AlCl3_low_anneal}(c). We also observe darker features similar to dimer vacancies, and features that are elongated across the dimer rows.  The chains themselves have begun to change in appearance with brighter segments visible within individual chains, as highlighted by the arrow in Fig.~\ref{AlCl3_low_anneal}(c).

\subsection{Investigating the Atomic Structure of the Chlorinated Aluminum Chains} \label{ssC}

To the authors' knowledge, the chains in Fig.~\ref{AlCl3_low_anneal} have not been observed before. This requires us to undertake a careful inspection of the chains to better understand the effect of annealing and the sequence of events leading up to incorporation. In this section, the chain makeup is considered, leading to a plausible atomic model.

\begin{figure*}
\includegraphics[width=\textwidth]{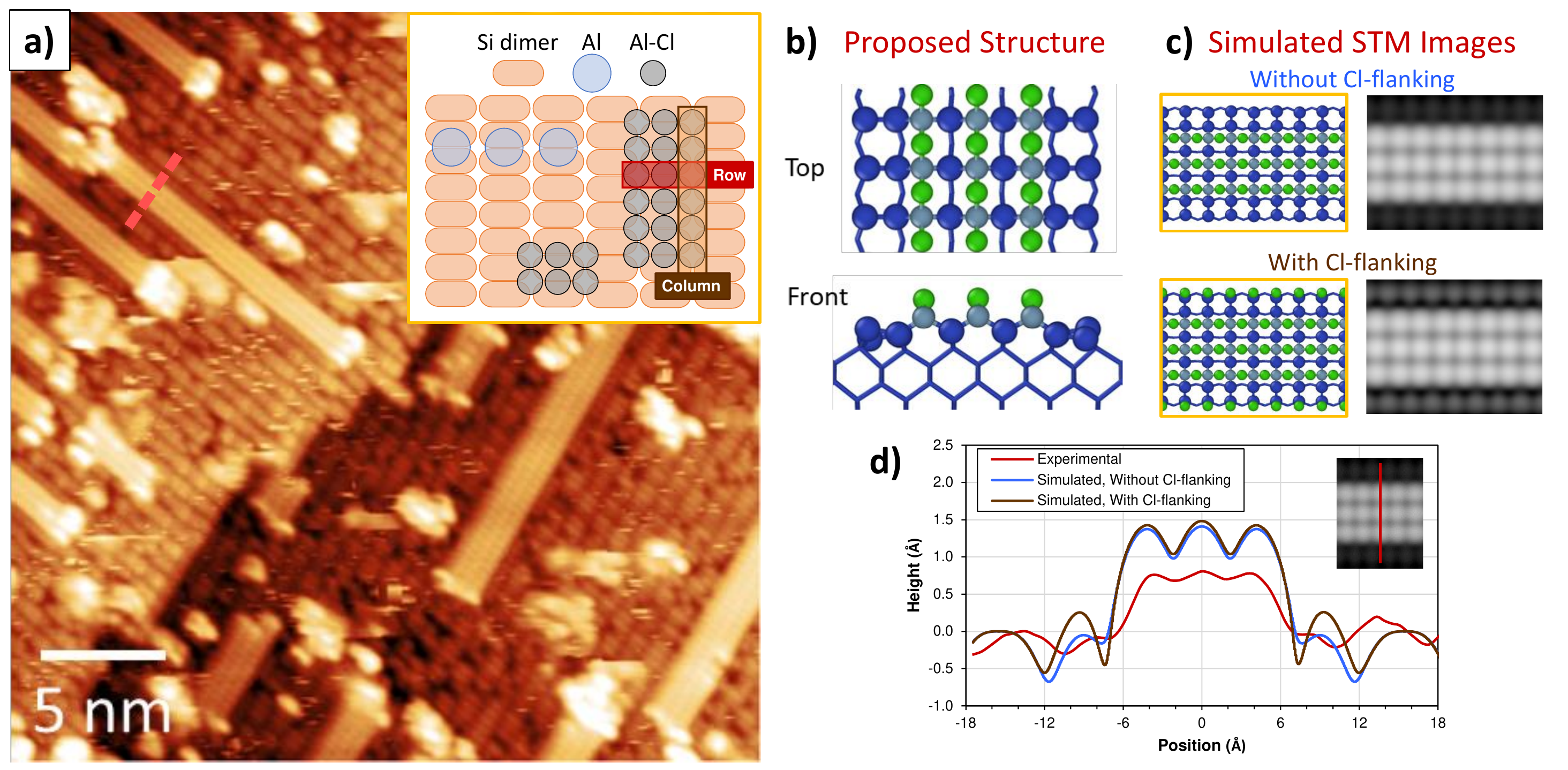}\hfill%
\caption{CAC analysis. (a) Filled-state STM image (-1.8 V, 0.25 nA) of AlCl$_{3}$-dosed Si(100) annealed to 400 $^{\circ}$C for 5 minutes. After annealing, long chains of chlorinated Al are seen on the surface running parallel to the Si dimer row direction. Chains are three spots wide and are centered on the dimer row bond as depicted by the schematic in the inset of (a). A typical Al chain is also shown. (b) The proposed atomic structure of the CAC. (c) Simulated filled-state (-1.8 V) STM images of the proposed CAC structure without and with Cl attached to the unbroken dimer rows alongside of the primary chain structure. (d) A line profile across one of the chains in (a) at the red line and the corresponding simulated line profiles. The simulation reproduces the three peaks of nearly equal height as seen experimentally. The experimental height of the chains is $\sim$0.7 \AA.}%
\label{3_atom_chain}%
\end{figure*}

The CACs are shown in a magnified image in Fig.~\ref{3_atom_chain}(a) and also shown schematically in the inset, where the repeat units are defined. We refer to a \textit{column} as a single linear sequence of features running parallel with the chain. A \textit{row} is used to denote the repeating unit spanning the entire width of the chain.  A closer look at the chains in Fig.~\ref{3_atom_chain}(a) reveals that the features consist of three adjacent columns with each row of three spots positioned in a line running between adjacent dimers in the same row, rather than the features being directly on or in line with a dimer. The middle unit is centered on the dimer row, whereas the side features fall roughly halfway between dimer rows. The CACs are shown schematically in the inset of Fig.~\ref{3_atom_chain}(a) for clarity. No wider or narrower chains of this three-column type were observed. Instead, side-by-side three-spot-wide chains can be found where the chains are spaced apart by an atomic site and no additional features are seen on the center of the intervening dimer row that would indicate the chains are linked tranversely. Moreover, we also observed that the chains abruptly end when they encounter one another on the surface. When this happens, the chains do not continue to grow past each other. These observations collectively suggest that the three-spot width is energetically favorable. In contrast to the CACs, a schematic of a typical Al chain running perpendicular to the dimer rows is also shown in the Fig.~\ref{3_atom_chain}(a) inset.\cite{itoh1993} For this Al chain, the STM spots should not be mistaken as the location of the Al atoms. Rather, the Al chain consists of Al dimers sitting between Si dimer rows. It is the Al-Si bonds that are most pronounced in the filled state, and for each Si dimer pair in the chain, the involved Al-Si bonds combine to produce the merged feature over the Si dimer pair.\cite{itoh1993,brocks1993}

Using the observations of the CAC features mentioned above and the clues gleaned from the STM simulations in Fig.~\ref{dft}, we propose a molecular structure for the CAC, which is shown in Fig.~\ref{3_atom_chain}(b). In the proposed chain structure, the middle dimer row is broken by Al atoms. The Al atoms sitting in the broken row are chained together by Cl atoms, since an Al-bonded Cl can form a second bond with a neighboring Al. In this way, the Cl atoms successfully terminate the Al bonds. Each Si atom in the broken dimer row is passivated by another Al, giving rise to two more AlCl columns, each sitting between dimer rows. It is important to note that this structure relieves the strain from the dimer reconstruction. The Si atoms in the broken dimer return to a near-perfect tetrahedral geometry.

We also consider a variant structure with additional Cl atoms on the neighboring dimer rows alongside of the chain. Both chains are shown Fig.~\ref{3_atom_chain}(c) alongside of the respective simulated STM images. The structures are labelled "Without Cl-flanking" and "With Cl-flanking". The energies can be compared by computing the average adsorption energy according to equation \ref{eq:chain}. The average adsorption energies with and without Cl-flanking is found to be -2.49 eV/AlCl\textsubscript{3} and -2.39 eV/AlCl\textsubscript{3}, respectively.

Overall, the simulated chain images shown in Fig.~\ref{3_atom_chain}(c) closely resemble the experimental images. Note that the Al atoms are not apparent in the simulated STM image. This is similar to the bridging AlCl in the D1 configuration (Fig.~\ref{dft}(b)). It is the Cl atoms sitting above the Al atoms that emerge in the image. Also, going along or across the chain, the spots are uniform in brightness, and this matches with the experimental image. While the addition of Cl on the chain flanks is a consideration, the effect on the STM image is subtle.

For additional detail, the simulated and experimental chain height profiles are compared in Fig.~\ref{3_atom_chain}(d). The simulated profile tracks the height along a linear path across the chain, as shown in the inset of Fig.~\ref{3_atom_chain}(d). The experimental height profile is taken along the dashed line going across the chain shown in Fig.~\ref{3_atom_chain}(a). The general features of the experimental profile are reproduced by the model. The simulated chain achieves the three peaks of nearly equal height, which is seen experimentally. Sources of uncertainty that could affect the apparent chain height comparison include the exact experimental tip-surface distance and the portion of the applied voltage being lost across the tunneling gap and contributing to surface band bending.\cite{bolotov2016} The difference in the location of the dimer peaks is attributed to the GGA functional used,\cite{perdew1996} as the model also predicts a slightly larger lattice constant than in experiment, consistent with trends in GGAs.\cite{kurth1999,ernzerhof1999} Consequently, this is expected to slightly widen the simulated chain also. Additionally, the finite size of the experimental tip is expected to smooth features in contrast with the point tip used in the calculations, leading to more defined features. The addition of Cl on the flanks only slightly alters the simulated profile, making it difficult to distinguish with certainty when comparing against experiment, especially considering that a realistic finite tip size prohibits fully imaging sharp dips.\cite{wang2004}

\begin{figure}
\centering
\includegraphics[scale=0.80]{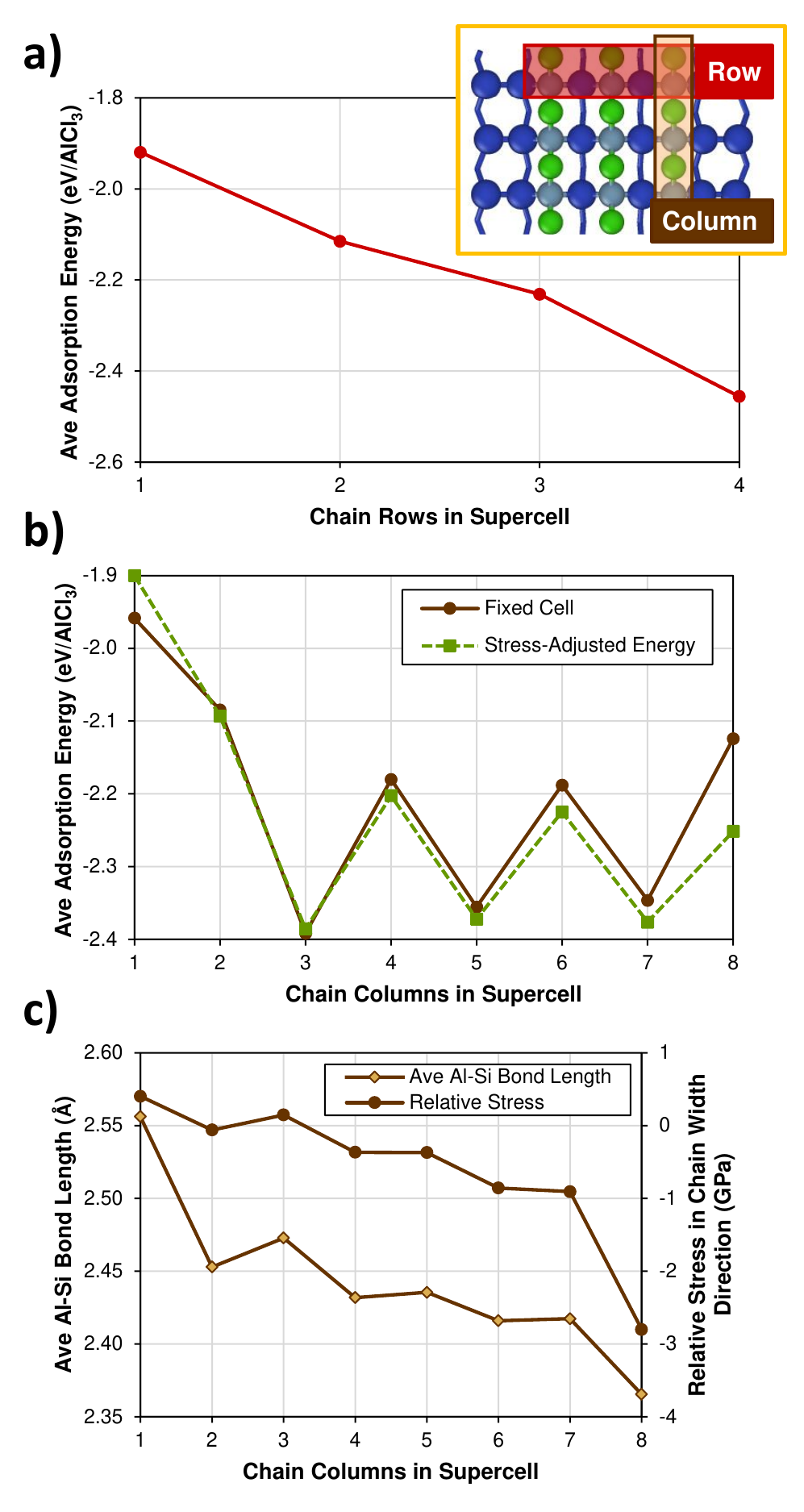}\hfill%
\caption{Chain length and width analysis of proposed CAC structure. The row and column repeating units of the chain are highlighted in the inset of (a). (a) The average adsorption energy as the number of rows is varied, demonstrating a preference for a long chain arrangement. (b) The average adsorption energy as the number of columns is varied, demonstrating the three-column width is most favorable. Results are shown for both fixing the cell dimensions and adjusting the cell dimensions to maintain constant stress. (c) Additional data on the fixed-cell model. The average Al-Si bond length and the supercell stress shows that the chain structure becomes compressed when more than three columns are present.}%
\label{dftchain}%
\end{figure}

For the proposed chain structure, variable width and variable length chains were computed to confirm that the structure energetically favors a long triple-wide chain. The repeat units are highlighted in Fig.~\ref{dftchain}(a) for clarity. In the present structure, a column is a single sequence of alternating Al and Cl atoms running parallel with the chain. A row consists of three Al atoms and three Cl atoms breaking one Si dimer as shown.

The chain length was varied using a supercell with two dimers and four dimers per row. The number of rows in the supercell was varied such that multiple rows were placed adjacently.  The model size permits up to four rows, and therefore, four rows represents an infinitely long chain due to periodic boundary conditions. Additionally, periodic boundaries mean that the one-, two-, and three-row models represent repeating short chains with the gap between chains decreasing as the chain is lengthened. Since a clear understanding of the chain ends has not been obtained at this point, no capping structure is included, and so it should be noted that the chain is simplified.

Fig.~\ref{dftchain}(a) shows the average adsorption energy as the number of rows in the supercell is varied. The average adsorption energy was computed by equation \ref{eq:chain}. When there is one row in the supercell the average adsorption energy is -1.92 eV/AlCl\textsubscript{3}. Upon increasing the number of rows, the average adsorption energy steadily improves. The average adsorption energy of the four-row model (infinite chain) is -2.46 eV/AlCl\textsubscript{3}, demonstrating that the long chain arrangement is more favorable than isolated rows or repeating short chains. This supports the possibility that the structure could form a chain. Ideally, chain formation can be asserted when the chain's energy can be compared with competitive arrangements. We have attempted an alternative chain with Al\textsubscript{2}Cl\textsubscript{3} rows reported in the Supplementary Information document. The average adsorption energy of the Fig.~\ref{3_atom_chain}(b) chain is superior to the Al\textsubscript{2}Cl\textsubscript{3} chain (and the adsorbates in Fig.~\ref{dft}). However, due to the guesswork in identifying competitive alternatives, we cannot assert conclusively the propensity for self-assembling into a chain. Nevertheless, the results in Fig.~\ref{dftchain}(a) suggest that chain formation is conceivable for the proposed CAC structure.

The number of columns in the chain construction was also investigated with a wider model, having four dimer rows and two dimers per row. This permits eight columns in the supercell, and therefore, having eight columns in the supercell represents an infinitely wide chain. For the other cases, the peridocity implies that the gap between chains decreases as the number of columns increases. All chains spanned the supercell in the chain's longitudinal direction and so were effectively infinite in length. We found that one column was not sufficient to break the dimer row. When a single-column chain was placed in a broken row, the dimers reformed upon geometric optimization, resulting in Si-Al-Si three-membered rings. At least two columns were needed for the broken row to be locally stable. Also, it was desirable to satisfy the dangling bonds from a broken dimer row before breaking other rows. Therefore, the variable-width chain was systematically configured so that the number of broken rows adheres to the following:
\begin{equation}
\text{Number of broken rows} = \left \lfloor{\frac{\text{Number of columns}}{2}}\right \rfloor
\end{equation}
The adsorption energy was computed according to equation \ref{eq:chain}. We consider two methods of handling the supercell, namely, fixing the cell dimensions and allowing for adjustments of the cell in the chain width direction. Due to hydrogen-termination of the bottomost layer while holding the bottom two Si layers to match the bulk lattice, the bare slab model registered slight compression. To account for this, the cell dimensions of the chain models were adjusted to maintain contant stress compared to the Si slab model.

The results of varying the chain width are shown in Fig.~\ref{dftchain}(b). Both the fixed-cell results and the stress-adjusted results are plotted. Aside from the single column chain, an odd number of columns is seen to be more favorable than an even number of columns, resulting in a sawtooth-like plot. This is because when there is an even number of columns, Si atoms from the broken dimer are left dangling on one side of the chain. More interestingly, the three-column width results in the lowest average adsorption energy, demonstrating it to be the most favorable width.

To explain the advantage of the triple-column construction, it is instructive to examine the fixed-cell models. In  Fig.~\ref{dftchain}(c), the average Al-Si bond length within the chain is plotted. As the chain becomes wider than three columns, the inner columns become overly compressed, evidenced by contracted Al-Si bonds. For comparison, the Al-Si bond lengths in the initial adsorbate structures were 2.44 \AA ~to 2.49 \AA. The normal stresses in the chain width direction, which are also plotted in Fig.~\ref{dftchain}(c), follow the trend in the Al-Si bond lengths. Note that the stress reported is the difference between the stresses in the chain model and the bare Si slab model.  Summarizing the results, the triple-column width is most favorable because all Si bonds in the broken dimer row are passivated and the Al-Si bonds are allowed sufficient space, introducing minimal stress. Again, in Fig.~\ref{dftchain}(c) we look to the fixed-cell models to gain a clear picture. When the cell dimensions are adjusted, the Al-Si bonds can somewhat relax leading to an improvement in the energies for the wider chains, but this is limited by the cost of expanding the Si layers.

Thus, the DFT results indicate that the proposed CAC structure demonstrates a preference for being in the form of an elongated triple-wide chain and is expected to image like the experimentally observed chain. The chain end structure remains to be studied, as well as the means for chain growth. Apart from these remaining points, the results presented here make a compelling case for the proposed structure.

Given the proposed atomic model, we can suggest that the self-assembling arises from AlCl fragments (produced by the D1 structure in Fig.~\ref{dft}) diffusing about the surface when activated by annealing. To support this idea, we compute the pathway for migrating a bridging AlCl fragment along the dimer row to the next bridging site when there are no other atoms in the vicinity of the AlCl fragment, and the energy barrier was found to be 1.05 eV. This indicates that AlCl migration should be activated by the annealing temperatures applied to the samples in Fig.~\ref{AlCl3_low_anneal}. The CAC appears to grow by adding a row at a time, because a single column does not extend far beyond the other columns (until chain break-up occurs). While the feature in Fig.~\ref{adsorption}(g) was observed and likely consists of a single column of the CAC, we do not find such long single-column chains in relation to the CACs. The ends appear to show a column advances by only one AlCl unit beyond an adjacent column in the same chain. This suggests there could be partial rows at the ends, and each partial end is a snapshot of some step along the chain growth process. Therefore, we suspect the chain growth occurs by thermally activated AlCl fragments diffusing about the surface and, on visiting a CAC end, bonding to the end, where it must be inclined to fill a row if there is a partial row or initiate a new row otherwise.

Additionally, with the proposed atomic model, we can better interpret the overlapping chains mentioned earlier and seen in the inset of Fig.~\ref{AlCl3_low_anneal}(a). We can identify this instance as having one column persisting and shared between the meeting chains. In this case, the interface of the two chain ends is five columns wide. Interestingly, while it possible to have one shared column between meeting offset chains, our model suggests it is unlikely to see two shared columns. This is because if the chains are offset by exactly one column, then one chain must be centered between dimer rows, which is unfavorable as it requires additional dimer breaking.

\subsection{Al Incorporation} \label{ssD}

\begin{figure*}
\centering
\includegraphics[width=\textwidth]{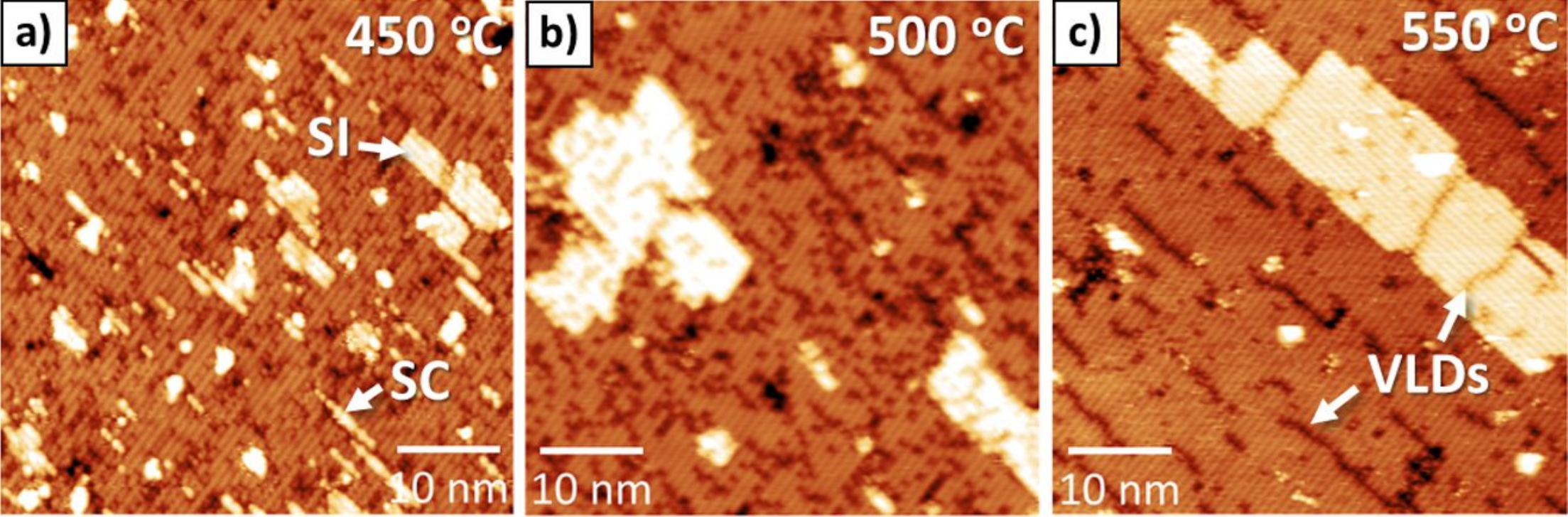}\hfill%
\setlength{\belowcaptionskip}{-7pt}
\caption{High-dosed samples, high-temperature annealing. Representative filled-state STM images (a: -1.85 V, 0.04 nA, b: -1.8 V, 0.15 nA, c: -1.8 V, 0.25 nA) of Si(100) dosed with 81 L of AlCl$_{3}$ and annealed for 5 minutes at (a) 450 $^{\circ}$C, (b) 500 $^{\circ}$C, and (c) 550 $^{\circ}$C. Islands (SI) and chains (SC) formed from ejected Si are visible due to the incorporation of Al atoms into the surface as well as Cl etching. In (a) and (b), Cl-terminated Si dimers are visible. In (c), vacancy line defects (VLDs) are formed due to the incorporated Al and can be observed on both the Si substrate and Si islands.}%
\label{AlCl3_high_anneal}%
\end{figure*}

Figure~\ref{AlCl3_high_anneal} shows filled-state STM images of Si(100) samples that were annealed to high temperatures, namely (a) 450 $^{\circ}$C, (b) 500 $^{\circ}$C, and (c) 550 $^{\circ}$C for 5 minutes. As previously, the AlCl\textsubscript{3} dosage is 81 L. Si chains and Si islands were observed on each surface (see the SC and SI marks in Fig.~\ref{AlCl3_high_anneal}) instead of the CACs. Recall that the CACs are oriented along the dimer row direction. In Fig.~\ref{AlCl3_high_anneal}(a), the dimer rows run from the lower left to the upper right, and clearly no chains are found resembling the CAC. We instead find different chains oriented perpendicular to the dimer row direction. These chains have a height of $\sim$0.13 nm and a width of $\sim$1.2 nm and are attributed to Si adatoms. In the vicinity to these chains, larger Si islands, having the same height as the Si chains, were observed. The additional Si features are most likely the result of Al atoms incorporating into the surface. At these elevated temperatures, the displaced Si atoms can easily diffuse about the surface and aggregate to form the observed chains and islands. This process has been observed and demonstrated for other dopant precursor molecules on Si(100), such as PH$_{3}$,\cite{lin2000} AsH$_{3}$,\cite{Curson:AsH3} and B$_{2}$H$_{6}$.\cite{Fuhrer:diborane} In addition to the Al incorporation process, we also expect Cl-involved etching processes to occur and contribute to the overall amount of the Si structures observed on the surface (see Supplemental Information).

The sizes of the Si islands increased with annealing at 500 $^{\circ}$C and 550 $^{\circ}$C. Also, the numbers of the islands decreased at the higher annealing temperatures. However, the total surface coverage of the islands showed no dependence on the annealing temperature (see Supporting Information). These findings suggest that the formation of larger islands is due to the merging of smaller islands and agglomerated Si chains, rather than additional Al being incorporated at higher temperatures.

There is also a noticeable evolution of the defects. At 450 $^{\circ}$C, many Cl-terminated Si dimer defects were observed, which will be further discussed in Part E. However, in Fig.~\ref{AlCl3_high_anneal}(b) and (c), we can observe vacancy line defect (VLD) structures on both the substrate and Si islands (see the VLDs in Fig.~\ref{AlCl3_high_anneal}). As suggested in the previous studies,\cite{zandvliet1995energetics} VLD structures are likely to be produced from Al incorporation into the Si substrate. At 550 $^{\circ}$C, the sample was overall less defective, which can be attributed to Al incorporation into the substrate as well as the loss of adsorbed Cl.

Estimating the amount of incorporated Al based solely on STM imaging has proven to be difficult. For the case of P, a 1-to-1 correspondence between incorporated P atoms and ejected Si atoms has generally been made after PH\textsubscript{3} exposure and annealing.\cite{lin2000} This enables a straight-forward determination of the amount of incorporated P simply by determining the amount of Si chains and islands observed in STM images. However, a direct comparison of the amounts of incorporated Al atoms to ejected Si atoms after AlCl$_{3}$ exposure is complicated by the presence of Cl on the surface, which is known to roughen and etch Si(100), producing both adsorbed Si islands and Si vacancies.\cite{herrmann2002,xu2003,butera2010adsorbate,de1998,agrawal2007} Therefore, we characterized the amount of incorporated Al with SIMS depth profiling.

\begin{figure}
\includegraphics[width=0.5\textwidth]{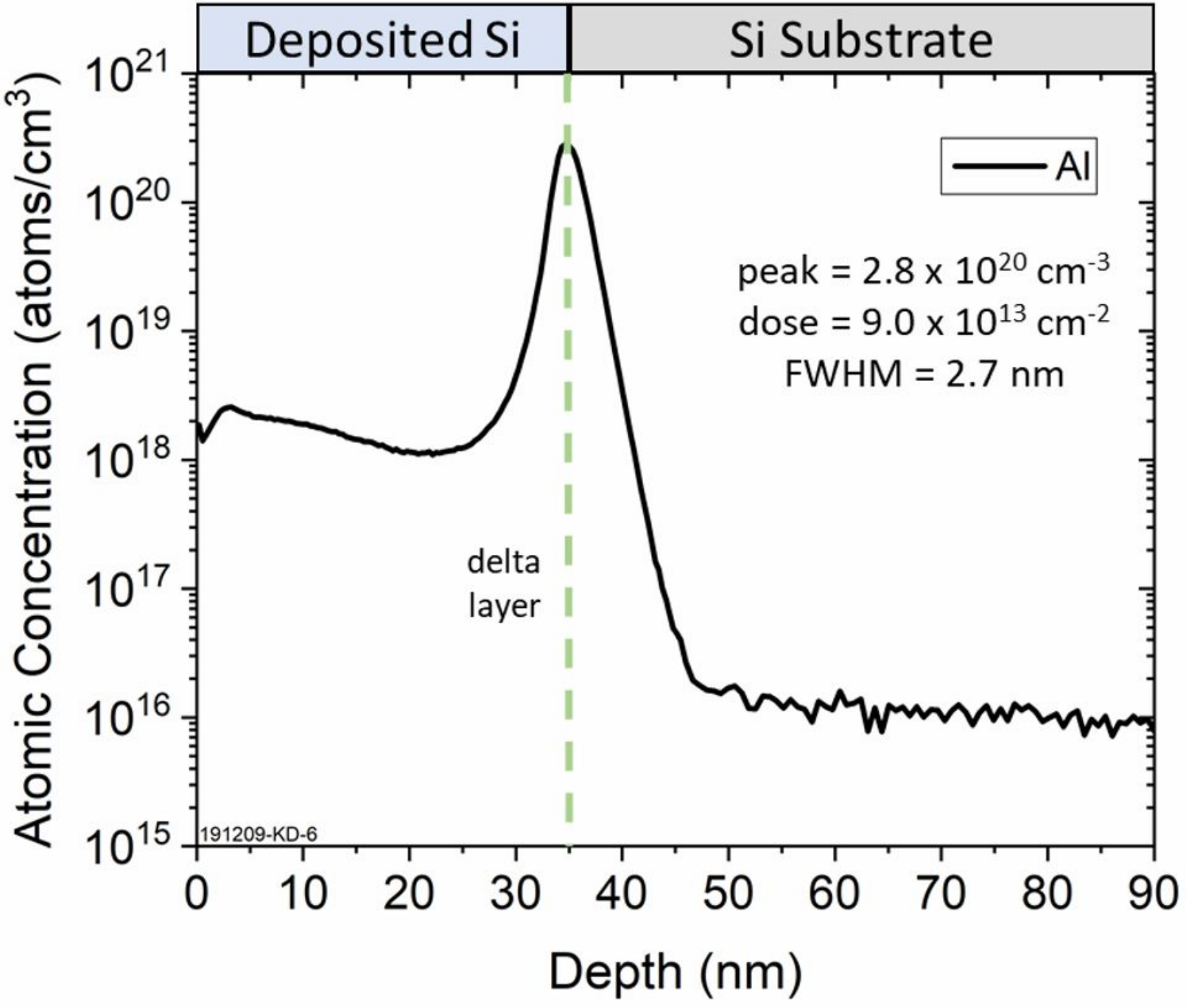}\hfill%
\caption{Secondary ion mass spectrometry Al depth profile of an Al $\delta$-layer formed from a 190 L AlCl$_{3}$ dose on Si(100) and capped with 35 nm of Si. The peak Al concentration is 2.8 $\times$ 10$^{20}$ cm$^{-3}$, the total Al density is 9.0 $\times$ 10$^{13}$ cm$^{-2}$, and the $\delta$-layer full width at half max is $\sim$2.7 nm. The Al content in the deposited film is due to Al contamination in the Si deposition source, and not Al diffusion from the $\delta$-layer.}%
\label{Al_SIMS}%
\end{figure}

After AlCl$_{3}$ exposure, samples were capped with $\sim$35 nm of Si to bury the Al, forming a $\delta$-doped layer and enabling SIMS depth profiling. Figure~\ref{Al_SIMS} shows a SIMS Al depth profile of a Si(100) sample that was dosed with $\sim$190 L AlCl$_{3}$. An incorporation anneal was not performed on this sample before Si capping to ensure Al was not lost through desorption. While the sample was not deliberately heated during Si deposition, radiant heat from the deposition source caused the sample to warm to $\sim$245 $^{\circ}$C. The peak Al concentration was found to be approximately 2.8 $\times$ 10$^{20}$ cm$^{-3}$ and was located 35 nm beneath the surface at the buried interface of the Al $\delta$-layer, as expected. We note that the Al content in the deposited film is due to some Al contamination in the Si deposition source (also observed by others\cite{SNL:Surface_Gate} using the same model) and not from AlCl$_{3}$ dosing, which occurs in a separate chamber. The total estimated areal dose of Al in this sample was 9.0 $\times$ 10$^{13}$ cm$^{-2}$. The upper bound of the Al $\delta$-layer thickness, as measured by the full width at half maximum (FWHM) of the peak, was found to be 2.7 nm.

\subsection{Cl Desorption} \label{ssE}

\begin{figure}
\includegraphics[width=0.5\textwidth]{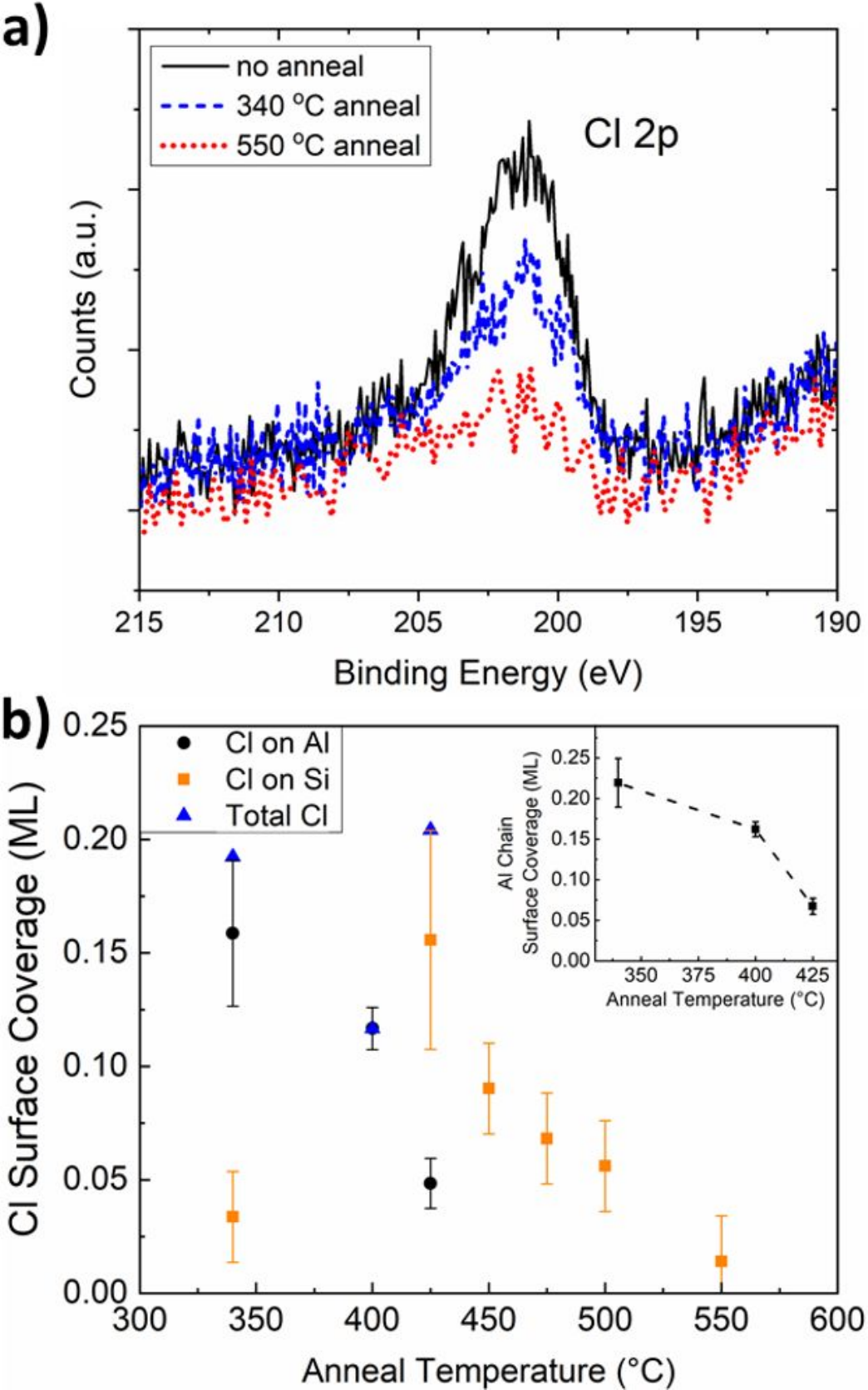}\hfill%
\caption{(a) X-ray photoelectron spectra of AlCl$_{3}$-dosed Si(100). The Cl 2p peak is shown immediately after dosing, after annealing to 340 $^{\circ}$C for 5 minutes, and after annealing to 550 $^{\circ}$C for 1 minute. The peak decreases with increasing temperature as Al-Cl species dissociate and Cl desorbs from the surface. (b) Shows the coverage of Cl-terminated Si dimers and Cl bound to Al (estimated from the proposed CAC structure) as a function of annealing temperature. Initially, Cl is locked into the CACs, producing few Cl-terminated dimers. Above 450 $^{\circ}$C, the CACs dissociated as Al is incorporated into the surface and Cl moved onto the surface forming terminated Si dimers. At higher temperatures, Cl begins to desorb and the concentration decreases again. }%
\label{XPS}%
\end{figure}

The use of a Cl-based dopant precursor presents a number of concerns for implementation within developed APAM process flows.  In particular, residual Cl is expected to negatively impact device performance. As a result, we monitored the Cl concentration on the sample through each processing step.  In-situ XPS was used to quantify the amount of Cl on the surface following exposure to AlCl$_{3}$. Figure~\ref{XPS}(a) shows the evolution of the Cl 2p peak from a sample that was exposed to 162 L of AlCl$_{3}$ at room temperature and then annealed at 340 $^{\circ}$C for 5 minutes, and then annealed further at 550 $^{\circ}$C for 1 minute.  The Cl 2p peak showed a steady decline in intensity as the sample was annealed providing a clear indication of Cl loss, which we presume to be through desorption of SiCl\textsubscript{2}.\cite{Gao:Cl_bonding,szabo1994}

As a comparison, Fig.~\ref{XPS}(b) plots the surface coverage of Cl as measured with STM after annealing the AlCl\textsubscript{3} exposed samples at temperatures between 340 $^{\circ}$C and 550 $^{\circ}$C. The data accounts for the Cl bound to surface Si dimers, which is easily identifiable by its distinctive darkened appearance,\cite{Chen:roughening} and Cl bound to Al, which is estimated from the proposed CAC model with 3 Cl atoms per unit length. Cl bound to Al was calculated from STM images where CACs were present and was a direct function of chain densities at the respective temperatures. The inset of Fig.~\ref{XPS}(b) shows the surface coverage of Al chains. The increase in Cl-terminated Si dimers observed from 340 $^{\circ}$C to 450 $^{\circ}$C is correlated with a decreased CAC surface coverage as shown in the inset of Fig.~\ref{XPS}(b). This observation implies that Cl was liberated from the chains and subsequently deposited onto the Si surface as part of the Al incorporation process as the annealing temperature was increased. Once adsorbed on Si dimers, the Cl was easily observed with STM. However, the total amount of Cl on the surface decreased linearly with temperature. The total coverage of Cl was most likely underestimated for the 400 $^{\circ}$C instance, where it was difficult to observe the Si terrace due to mobile species on the surface, and overestimated for 425 $^{\circ}$C where CACs may not be fully covered with Cl. Recall that in Fig.~\ref{AlCl3_low_anneal}(c) the appearance of the CACs was altered at this temperature and may be an indication of reduced Cl content within the chains. The fact that the increased concentration of Cl bound to Si was correlated with the decreased concentration of chains, combined with the Cl 2p XPS signal shown in Fig.~\ref{XPS}(a), provides further experimental validation of the presence of Cl within the chains and the proposed structure given in Fig.~\ref{3_atom_chain}(b).

Further annealing beyond 450 $^{\circ}$C led to Al incorporation into the Si surface, as shown in Fig.~\ref{AlCl3_high_anneal}, and a drop in both the Cl 2p XPS signal and the STM quantified Cl surface coverage.  By 550 $^{\circ}$C most of the Cl has been removed from the surface. It should be noted that neither Cl nor Cl\textsubscript{2} are expected to desorb on their own, but Cl etching mechanisms are expected to be active near the temperatures studied here.  It has been previously shown that SiCl\textsubscript{2} has a peak desorption temperature as low as $\sim$575 $^{\circ}$C, depending on the sample dosing conditions, \cite{Gao:Cl_bonding} although it is more commonly found to occur closer to 675 $^{\circ}$C.\cite{CL_etch:Dohnalek, MENDICINO:Cl_adsorption} High doping concentrations are also known to lower the peak desorption temperature of SiCl\textsubscript{2} \cite{Gates:H_Cl_exchange} and it is not surprising to observe a substantial loss of Cl after annealing at 550 $^{\circ}$C.

\section{Conclusion} \label{conclusion}

We have demonstrated the viability of AlCl$_{3}$ as a molecular precursor for the incorporation of Al on Si(100) with the goal of furthering process development for atomically precise acceptor-based devices. The feasibility of AlCl$_{3}$ is in part due to its ease in adsorbing to Si, evidenced by STM imaging of room temperature-dosed samples and by DFT calculations demonstrating barrierless adsorption.   AlCl$_{3}$ is also a promising candidate because Al incorporation was evidenced at about 450 $^{\circ}$C to 500 $^{\circ}$C. This temperature is not unreasonably high, but it is near the peak thermal desorption temperature of H$_{2}$ (520 $^{\circ}$C to 530 $^{\circ}$C) from Si(100).\cite{wise1991} While it remains to be proven, it is expected that the incorporation anneal for Al from AlCl\textsubscript{3} will invalidate the use of a hydrogen based resist. In contrast, Cl appears to be a more compatible resist, since Cl is mainly removed as SiCl\textsubscript{2} (rather than Cl\textsubscript{2}) with a peak thermal desorption temperature of about 700 $^{\circ}$C.\cite{szabo1994}

Our investigation led to the further discovery of unique self-assembled CACs. An atomic structure of the CAC was proposed with supporting DFT calculations. The formation of the CACs reveals AlCl diffusion and likely plays a critical role in the final arrangement of the incorporated Al, if the Al atoms directly incorporate from their positions within the chain. It remains to be seen whether the self-assembling behavior will present significant difficulties in device fabrication. To further understand these effects, a determination of the electrical activity of the $\delta$-layers will be made. Even if the CAC formation proves to be detrimental, the use of a patterned resist ought to inhibit AlCl diffusion, as the adsorbed AlCl species are expected to be trapped within their respective windows. Further avenues of study are made available by the discovery of the CACs. The following points remain unsolved and warrant future work: the steps in CAC self-assembly, the CAC end structure, the steps in CAC break-up and Al incorporation from the chain, and the CAC's role in the final distribution of incorporated Al atoms. And apart from APAM applications, the self-assembling behavior of dissociated AlCl$_{3}$ adsorbates may have ramifications on the morphology of films prepared by atomic layer deposition using AlCl$_{3}$, which may be another opportunity for investigation.

In order to fabricate atomically precise acceptor devices, comprehensive knowledge of the adsorption and dissociation of the precursor, the diffusion and self-assembly of the resulting species, and incorporation of the acceptor dopant is unquestionably critical. As previously demonstrated for phosphine, a specific configuration of adsorbates is required to produce deterministic incorporation of a single phosphorus atom.\cite{Fuechsle_Atom:2012} Similarly, to deterministically drive a single acceptor atom to incorporate into Si necessitates deep insight into what adsorbate configurations and conditions naturally lead to the desired incorporation. Simply put, manipulation requires mastery, and this work is a step towards mastering the complexities of AlCl\textsubscript{3} on Si.

\section{Acknowledgement}

The authors acknowledge the computational facilities from the University of Maryland supercomputing resources and the Maryland Advanced Research Computing Center (MARCC). This work was supported in part by the Laboratory Directed Research and Development program at Sandia National Laboratories, a multimission laboratory managed and operated by National Technology and Engineering Solutions of Sandia, LLC., a wholly owned subsidiary of Honeywell International, Inc., for the U.S. Department of Energy's National Nuclear Security Administration under contract DE-NA-0003525. This paper describes objective technical results and analysis. Any subjective views or opinions that might be expressed in the paper do not necessarily represent the views of the U.S. Department of Energy or the United States Government.

\section{Supporting Info}

Description and comparison DFT modeling setups. Supplementary images of adsorbate structures. STM image of HCl contamination when dosing without Ar. Adsorption and dissociation pathway for Al\textsubscript{2}Cl\textsubscript{6} on Si(100). Al\textsubscript{2}Cl\textsubscript{3} chain analysis. STM image of Cl etching on Si(100). Si island statistics for samples annealed at temperatures 450-550 $^{\circ}$C.

\bibliography{AlCl3_incorporation_JPCC}

\end{document}